\documentclass[10pt]{article}
\usepackage{amsmath,amssymb,amsfonts}
\usepackage{units}
\usepackage{bbold,euler}
\usepackage{graphicx}
\usepackage{dcolumn}
\usepackage{bm,bbm,mathtools,esvect}
\usepackage{slashed}
\usepackage{esvect}
\usepackage{fullpage}
\usepackage[utf8]{inputenc}
\title{\sc Spin and angular momentum in\\  quaternionic quantum mechanics }
\author{{\sf SERGIO GIARDINO\footnote{\tt sergio.giardino@ufrgs.br}}\\
\\
\small \it Departamento de Matem\'atica Pura e Aplicada \\
\small \it Universidade Federal do Rio Grande do Sul (UFRGS)\\
\small \it Caixa Postal 15080, 91501-970  Porto Alegre RS \\
\small \it Brazil}
\begin{document}
\date{} 
\maketitle

\begin{abstract}
\noindent We present two novel solutions of real Hilbert state quaternionic quantum mechanics ($\mathbbm H$QM). Firstly, we observe that the angular momentum operator admits two different classes of physically non-equivalent free particles. As a second result, we study the Larmor precession to observe that it has a quaternionic solution where a novel phenomenological interpretation is possible, as well as a different of spin is possible, and these results may encourage experimental and theoretical investigations of the quaternionic theory.

\vspace{2mm}
\noindent keywords: quantum mechanics; formalism; quantized spin models.

\vspace{1mm}
\noindent pacs numbers: 03.65.-w; 03.65.Ca; 75.10.Jm.
\end{abstract}

\section{\;\sc Introduction\label{I}}

In this note we report a simple result in quaternionic quantum mechanics ($\mathbbm{H}$QM). This quantum theory intends to
generalize quantum mechanics by replacing the complex wave functions of the usual quantum mechanics ($\mathbbm{C}$QM) with
quaternionic functions. We only mention that  quaternionic numbers ($\mathbbm{H}$) generalize complex numbers
by consistently adding two complex units to complex numbers. In general, if $q\in\mathbbm{H}$, then
\begin{equation}
q=x_0+x_1\,i+x_2\,j+x_3\,k,\qquad\mbox{and}\qquad i^2=j^2=k^2=-1.
\end{equation}
The quaternionic complex units $i,\,j$ and $k$ are endowed with a global anti-symmetric property, such that $ij=-ji$ and so on for the remaining complex units. There is of course a whole theory concerning quaternions, and we mention \cite{Ward:1997qcn,Kramer:2017fnn} as nice introductions to
the subject. The application of quaternions to quantum mechanics has their first historical moments mainly subsumed
in Stephen Adler's book \cite{Adler:1995qqm}, and this theory is built using anti-hermitian hamiltonian operators. A particularly serious drawback
of this theory is their ill-defined classical limit ({\it c.f.} sec. 4.4 of \cite{Adler:1995qqm}). Moreover, it is usually hard to find simple solutions of this theory whose physical interpretation is also accordingly very puzzling. Several examples of this situation can be mentioned \cite{Davies:1989zza,Davies:1992oqq,Ducati:2001qo,Nishi:2002qd,DeLeo:2005bs,Madureira:2006qps,Ducati:2007wp,Davies:1990pm,DeLeo:2013xfa,DeLeo:2015hza,Giardino:2015iia,Sobhani:2016qdp,Procopio:2016qqq,Sobhani:2017yee,DeLeo:2019bcw,Hassanabadi:2017wrt,Hassanabadi:2017jiz}.  

This situation of the anti-hermitian $\mathbbm H$QM exhibits a strong contrast to $\mathbbm{C}$QM, where one can resort to several simple solutions in order to have physical insight. Obvious examples are the square well, the hamonic oscillator, the hydrogen atom, and so on.
Consequently, the outstanding question of the physical interpretation of $\mathbbm{H}$QM naturally requires a mathematical framework capable of finding exact solutions. A solution to this problem has been found removing the anti-hermiticity condition of the Hamiltonian operator of the Schroedinger operator \cite{Giardino:2018lem}, and imposing a real Hilbert space \cite{Giardino:2022kxk}. These changes implies the $\mathbbm{H}$QM in the real Hilbert space as a more consistent theory, where the classical limit, and the Ehrenfest theorem also hold. Several exact solutions could be found within this novel mathematical and conceptual framework, namely the free particle \cite{Giardino:2017yke,Giardino:2017pqq}, the Virial theorem, and the quantum Lorentz solution \cite{Giardino:2019xwm}, the square well potentials \cite{Giardino:2020cee}, the harmonic oscillator \cite{Giardino:2021ofo}, and the quantum scattering \cite{Giardino:2020ztf}. Furthermore, relativistic solutions were also found, namely the Klein-Gordon \cite{Giardino:2021lov}, the Dirac solution \cite{Giardino:2021mjj}, the scalar field \cite{Giardino:2022kxk} and the Dirac field \cite{Giardino:2022gqn}.

These former results inspire us to research further exact solutions of $\mathbbm H$QM in the real Hilbert space to reinforce our hypothesis that this theory is the correct formulation of a quaternionic quantum theory. In this paper, another exact solution is presented, this time  quaternionic solution of the electron in a magnetic field. This solution appear early in quantum mechanics \cite{Dirac:1937pqm}, but it was unknown in the quaternionic theory, and therefore it is important to investigate whether the quaternionic and the complex quantum formulations have a common phenomenological background that hold them. In order to solve this problem, we also investigated the quaternionic momentum operator, and observed that two physically non-equivalent quaternionic eigenfunctions to the momentum operator are possible. The positive answer to these questions encourage us to seek the physical differences that comprise exciting directions of future research. However, we must remember that there are previous studies concerning quaternionic models of $\mathbbm{C}$QM involving spin, for example \cite{Cahay:2019bqp}, but these solutions are comprised within the $\mathbbm C$QM context.

The article is organized in three sections. We firstly discuss in the following the angular momentum in general and show that the quaternionic theory seems very similar to the complex case. In Section \ref{S} we 
show that both of the theories may have different prediction to the usual case of the Larmor precession of  the spin angular momentum.
We will see that both theories may coincide in precise of certain parameters, something that was never achieved using anti-hermitian
quaternionic theory. A conclusion section rounds off this article with our interpretation of the results and directions for future research.

\section{\;\sc Angular momentum\label{A}}
Before considering the angular momentum operator, let us consider the
quaternionic differential equation
\begin{equation}\label{a01}
\frac{d\Lambda}{dx}=\mu\Lambda,
\end{equation}
where $\,\Lambda(x)\,$ is a unit quaternionic function and $\,\mu\,$ is a quaternionic constant. We shall consider $\,\Lambda\mu\,$ as
a possibility to the right hand side of (\ref{a01}) as well. Let us entertain  three possibilities for unit quaternions in (\ref{a01}):
\begin{equation}\label{a02}
\Lambda_1=\cos\theta_0\, e^{imx}+\sin\theta_0\,e^{imx}j,\qquad\Lambda_2=\cos\theta_0\, e^{imx}+\sin\theta_0\,e^{-imx}j,\qquad
\Lambda_3=\cos mx\, e^{i\Gamma_0}+\sin mx\,e^{i\Omega_0}j.
\end{equation}
The unit quaternions (\ref{a02}) in (\ref{a01}) give
\begin{equation}\label{a03}
\frac{d\Lambda_1}{dx}\,=\,im\Lambda_1,\qquad\qquad\frac{d\Lambda_2}{dx}\,=\,\Lambda_2\,im,\qquad\qquad
\frac{d\Lambda_3}{dx}\,=\,m e^{i(\Gamma_0+\Omega_0)}j\,\Lambda_3\,=\,\Lambda_3\,	m\,j\, e^{i(\Gamma_0-\Omega_0)}.
\end{equation}
The quaternionic constant $\,\mu\,$ is pure imaginary in every case, and therefore 
\begin{equation}\label{a04}
\frac{d^2\Lambda}{dx^2}=-m^2\Lambda
\end{equation}
holds in the three cases. Considering the linear momentum as defined in \cite{Giardino:2018lem}, 
\begin{equation}\label{a05}
\bm p\Psi=-\hbar\bm\nabla\Psi i,
\end{equation}
we introduce the notation
\begin{equation}\label{a06}
\big(-\hbar\bm\nabla\big| i\big)\Psi=-\hbar\bm\nabla\Psi i.
\end{equation}
Equations (\ref{a03}) indicate that  $\,\Lambda_1\,$ admits left eigenvalue after the action of the derivative operator on it, $\,\Lambda_2\,$ admits a right eigenvalue, and $\,\Lambda_3\,$  admits eigenvalues on both of their sides. The three functions can be eigenfunctions of momentum, and this 
is expected because all of them are interpreted as either free particles \cite{Giardino:2017yke,Giardino:2017pqq} or confined particles \cite{Giardino:2020cee}. Using the usual definition of the angular momentum
\begin{equation}\label{a07}
L_a=\epsilon_{abc}x_bp_c,\qquad\mbox{where}\qquad a,\,b,\,c=\{1,\,2,\,3\},
\end{equation}
where $\epsilon_{abc}$ is the anti-symmetric Levi-Civit\`a symbool, the angular momentum can be generalized for a imaginary constant unit $\,\eta,\,$ so that
\begin{equation}\label{a08}
\eta^2=-1,\qquad\mbox{ and either}\qquad \eta=i\qquad\mbox{ or}\qquad\eta=j\, e^{i(\Gamma_0-\Omega_0)}.
\end{equation}
Irrespective of the imaginary unit, the usual commutation relations hold, 
\begin{equation}\label{a09}
\qquad\big[L_a,\,L_b\big]=\hbar\epsilon_{abc}\big(L_c|\eta\big),\qquad \big[L^2,\,\bm L\big]=0,
\end{equation}
and additionally the total angular momentum $L^2$ has eigenvalues $\,\hbar\ell(\ell+1)\,$ where $\ell$ is an integer or a half-integer and $\,L_3\,$ has eigenvalue $\,m=\{-\ell,\cdots ,\,\ell\}\,$.  In polar coordinates $(r,\,\theta,\,\phi)$, we choose 
\begin{equation}\label{a10}
L_3=\hbar\left(\frac{\partial}{\partial\phi}\Big|\eta\right),
\end{equation} 
and equations (\ref{a02}) are identical to eigen-value equations for $L_3$. The equations for in radial and $\theta$ direction does not change because $L^2$ is a real operator. Thus we conclude that the eigenfunctions for $L^2$ and $L_3$ are the quaternionic spherical harmonics discussed in the harmonic oscillator \cite{Giardino:2021ofo}, namely
\begin{equation}\label{a11}
\mathcal{Y}_\ell^m(\theta,\,\phi)=\sigma\sqrt{\frac{(2\ell+1)}{4\pi}\,\frac{(\ell-|m|)!}{(\ell+m)!}}\,P^m_\ell(\cos\theta)\,\Lambda(\phi).
\end{equation}
$P^m_\ell(\cos\theta)$ are associate Legendre polinomials while $\Lambda$ is either $\Lambda_2$ or $\Lambda_3$. This is the first angular momentum solution  to the quaternionic Schr\"odinger equation, and
the quaternionic angular momentum and the complex result angular momentum have identical expectation value. On the other hand, imposing $\Lambda=\Lambda_3$ implies a change in the imaginary unit in the definition of the angular momentum operator. 

A conceptual bonus from the results is the interpretation of the imaginary unit in quantum mechanics as a symmetry operator that determines
the unitary quaternions comprising the wave functions. We have different options for $\eta$ depending of 
the case. In general, we may write the Schr\"odinger equation for a Hamiltonian $\mathcal{H}$
\begin{equation}\label{a12}
\hbar\frac{\partial\Psi}{\partial t}\eta=\mathcal{H}\Psi.
\end{equation}
Although very simple, this is a novel result, and indicate a way that can be useful in order to simplify problems and to choose the physically relevant solutions of a problem. Now we entertain a situation where the quaternionic description in fact gives a result that is different from the complex case.
\section{\sc Quaternionic spin and the Larmor precession\label{S}}
From the angular momentum discussion, we infer that the spin algebra in quaternionic quantum mechanics is identical to the complex case
\begin{equation}\label{s01}
\qquad\big[S_a,\,S_b\big]=\hbar\epsilon_{abc}\big(S_c|\eta\big),\qquad \big[S^2,\,\bm S\big]=0,
\end{equation}
and of course the eigen-values equations are
\begin{equation}\label{s02}
S^2\Psi_{sm}=\hbar^2s(s+1)\Psi_{sm},\qquad S_3\Psi_{sm}=\hbar m\Psi_{sm}
\end{equation}
The $s=1/2$ case is such that $\bm S=\hbar\bm\sigma/2$, where $\sigma_a$ are the Pauli spin matrices. Let us entertain two examples of quaternionic spin where there are differences between the complex and quaternionic results.
The Hamiltonian of a spinning electron at rest of charge $e$ and mass $m$ is
\begin{equation}\label{s03}
\mathcal{H}=-\gamma\bm{B\cdot S}
\end{equation}
where $\gamma=e/2m$ is the gyromagnectic ratio of the electron and $\bm B$ is the magnetic field. Supposing an uniform magnetic 
field $\bm B=B_0\hat{\bm x}_3$, the Hamiltonian of an electron of energy $E$ will be
\begin{equation}\label{s04}
\mathcal{H}=-E\sigma_3\qquad\mbox{where}\qquad E=\frac{\hbar\gamma B_0}{2}.
\end{equation}
There are two cases to be considered.
\subsection{$\eta=i$}
The solution for (\ref{a12}) with $\eta=i$ using the eigenfunctions of (\ref{a10}), also with $\eta=i$, accordingly reads
\begin{equation}\label{s05}
\Psi=\cos\theta
\left[
\begin{array}{c}
\cos\left(\frac{\alpha}{2}\right) e^{i\frac{E}{\hbar}t}\\
\sin\left(\frac{\alpha}{2}\right) e^{-i\frac{E}{\hbar}t}
\end{array}
\right]
+
\sin\theta
\left[
\begin{array}{c}
\cos\left(\frac{\beta}{2}\right) e^{-i\frac{E}{\hbar}t}\\
\sin\left(\frac{\beta}{2}\right) e^{i\frac{E}{\hbar}t}
\end{array}
\right]j
\end{equation}
Where $\alpha$,  $\beta$, and $\theta$ were chosen as normalization parameters for the wave function. As expected, $\theta=0$ recovers the usual complex result.
The expectation values 
\begin{equation}
\langle S_a\rangle=\mathfrak{Re}\Big[\Psi^\dagger S_a \Psi\Big]
\end{equation}
are so that
\begin{align}\label{s06}
&\big\langle S_1\big\rangle=\frac{\hbar}{2}\big(\cos^2\theta\sin\alpha+\sin^2\theta\sin\beta\big)\cos\omega t\\
&\big\langle S_2\big\rangle=\frac{\hbar}{2}\big(\cos^2\theta\sin\alpha+\sin^2\theta\sin\beta\big)\sin\omega t\\
&\big\langle S_3\big\rangle=\frac{\hbar}{2}\big(\cos^2\theta\cos\alpha+\sin^2\theta\cos\beta\big)
\end{align}
where $\,\omega=-\gamma B_0\,$ is the Larmor precession. 
In agreement to the classical result,
the spin vector precesses around the field vector with angular velocity $\omega$. However, differently to the complex quantum case, the spin vector is splitted between a complex spin vector and a quaternionic spin vector. The complex vector performs an angle $\alpha$ with $\bm B$ and the quaternionic vector performs an angle $\beta$ with $\bm B$. The angular velocity of precession is identical in both of the cases, and the angle between them is of course $\alpha-\beta$. 
If $\alpha=\beta$, we recover the complex result, where the spin vector and the magnetic field perform an angle $\alpha$, and thus 
the quaternionic case is more general. The research of physical situations where $\,\alpha\neq\beta\,$ are natural directions for future investigation.
We get an interesting insight of this result from
\begin{equation}\label{s07}
\sum_{a=1}^3\big\langle S_a^2\big\rangle=\frac{\hbar^2}{4},\qquad\qquad
\sum_{a=1}^3\big\langle S_a\big\rangle^2=\frac{\hbar^2}{4}\Big[\cos^4\theta+\sin^4\theta+2\sin^2\theta\cos^2\theta\cos\big(\alpha-\beta\big)\Big]
\end{equation}
and consequently the standard deviation is
\begin{equation}\label{s08}
\sigma_S=\frac{\hbar}{2}\left|\sin 2\theta\sin\frac{\alpha-\beta}{2}\right|.
\end{equation}
We recover the complex result if $\theta=0$ or if $\alpha=\beta$. This indicates that quaternionic and quantum mechanics predict identical
or different results depending on the parameters. The parameter $\theta$ indicates a deviation from the complex wave function, while $\alpha$ and $\beta$ are geometric parameters concerning the precession of the dipole moment. The quaternionic case has two precessions, and $\alpha=\beta$ situation is particularly curious because the quaternionic and complex theory have identical expectation values, because the precession is identical in both of the components, the  pure complex and the pure quaternionic. Finally, we observe that this result indicate possibly the simplest way to a phenomenological 
test of $\mathbbm H$QM, where (\ref{s08}) is of course the quantity to be experimentally tested. Let us consider the second possibility.

\subsection{$\eta=e^{i(\alpha-\beta)}j$}

In this case, the normalized wave function will be
\begin{equation}\label{s05}
\Psi=
\frac{e^{i\alpha}}{\sqrt{2}}\left[
\begin{array}{c}
\cos\left(\frac{\omega}{2}t\right) \\ \\
\sin\left(\frac{\omega}{2}t\right) 
\end{array}
\right]
+
\frac{e^{i\beta}}{\sqrt{2}}\left[
\begin{array}{c}
-\sin\left(\frac{\omega}{2}t\right) \\ \\
\cos\left(\frac{\omega}{2}t\right) 
\end{array}
\right]j
\end{equation}
Accordingly,
\begin{eqnarray}\label{s10}
\big\langle S_1\big\rangle=\big\langle S_2\big\rangle=\big\langle S_3\big\rangle=0,
\end{eqnarray}
but
\begin{equation}
 \big\langle\bm S^2\big\rangle=\frac{3}{4}\hbar^2.
\end{equation}
From the standpoint of $\mathbbm C$QM, this result is completely nonsense. It is simply impossible. However, from the standpoint of $\mathbbm H$QM, we have a remarkable novel result. It indicates that a fundamentally different quaternionic spin, whose theory is still to be built. And this is of course an exciting direction of future research.

\section{\sc Conclusion \label{C}      }

In this letter we present two simple and novel results concerning $\mathbbm H$QM in the real Hilbert space. The results of the second section of the paper enabled us to generalize the quaternionic momentum operator and identify two non-equivalent classes of quaternionic free particles. Finally, in the third section of the paper, we studied the Larmor precession in $\mathbbm H$QM, and observed that the quaternionic case may have a phenomenological significance, a novel and simple result that may encourage the experimental study. Moreover, we observed that a fundamentally different feature of spin in $\mathbbm H$QM indicate that this concept is not equivalent in both of the theories.

The  directions for future research are very interesting, we mention the quaternionic Stern-Gerlach study as an interesting possibility, as well as the addition of angular momenta, a generalization of spin, or even quaternionic Hopf algebras. In fact, an exciting future for $\mathbbm H$QM research can to predicted.

%
%
%
%

\bibliographystyle{unsrt} 

\begin{thebibliography}{10}

\bibitem{Ward:1997qcn}
\texttt{J. P. Ward}.
\newblock {\em {``Quaternions and Cayley Numbers''}}.
\newblock {\it Springer Dordrecht} (1997).

\bibitem{Kramer:2017fnn}
\texttt{J. Kramer; A.-M. von Pippich}.
\newblock {\em {``From natural numbers to quaternions''}}.
\newblock {\it Springer Dordrecht} (2017).

\bibitem{Adler:1995qqm}
\texttt{S. L. Adler}.
\newblock {``Quaternionic Quantum Mechanics and Quantum Fields''}.
\newblock Oxford University Press (1995).

\bibitem{Davies:1989zza}
{\tt A. J. Davies; B. H. J. McKellar}.
\newblock {``Nonrelativistic quaternionic quantum mechanics in one
  dimension''}.
\newblock {\em Phys. Rev.}, {\bf A40}:4209--4214, (1989).

\bibitem{Davies:1992oqq}
{\tt A. J. Davies; B. H. J. McKellar}.
\newblock {``Observability of quaternionic quantum mechanics''}.
\newblock {\em Phys. Rev.}, {\bf A46}:3671--3675, (1989).

\bibitem{Ducati:2001qo}
\texttt{S. De Leo; G. Ducati}.
\newblock {`` Quaternionic differential operators''}.
\newblock {\em J. Math. Phys}, {\bf 42}:2236--2265, (2001).

\bibitem{Nishi:2002qd}
\texttt{S. De Leo; G. Ducati; C. Nishi}.
\newblock {`` Quaternionic potentials in non-relativistic quantum mechanics''}.
\newblock {\em J. Phys}, {\bf A35}:5411--5426, (2002).

\bibitem{DeLeo:2005bs}
\texttt{S. De Leo; G. Ducati}.
\newblock {`` Quaternionic bound states''}.
\newblock {\em J. Phys}, {\bf A35}:3443--3454, (2005).

\bibitem{Madureira:2006qps}
\texttt{S. De Leo; G. Ducati; T. Madureira}.
\newblock {``Analytic plane wave solutions for the quaternionic potential
  step''}.
\newblock {\em J. Math. Phys}, {\bf 47}:082106--15, (2006).

\bibitem{Ducati:2007wp}
\texttt{S. De Leo; G. Ducati}.
\newblock {`` Quaternionic wave packets''}.
\newblock {\em J. Math. Phys}, {\bf 48}:052111--10, (2007).

\bibitem{Davies:1990pm}
\texttt{A. J. Davies}.
\newblock {``Quaternionic Dirac equation''}.
\newblock {\em Phys.Rev.}, {\bf D41}:2628--2630, (1990).

\bibitem{DeLeo:2013xfa}
{\tt S. De Leo; S. Giardino}.
\newblock {``Dirac solutions for quaternionic potentials''}.
\newblock {\em J. Math. Phys.}, {\bf 55}:022301, (2014).

\bibitem{DeLeo:2015hza}
{\tt S. De Leo; G. Ducati; S. Giardino}.
\newblock {``Quaternioninc Dirac Scattering''}.
\newblock {\em J. Phys. Math.}, {\bf 6}:1000130, (2015).

\bibitem{Giardino:2015iia}
{\tt S. Giardino}.
\newblock {``Quaternionic particle in a relativistic box''}.
\newblock {\em Found. Phys.}, {\bf 46}(4):473--483, (2016).

\bibitem{Sobhani:2016qdp}
{\tt H. Sobhani; H. Hassanabadi}.
\newblock {``Scattering in quantum mechanics under quaternionic Dirac delta
  potential''}.
\newblock {\em Can. J. Phys.}, {\bf 94}:262--266, (2016).

\bibitem{Procopio:2016qqq}
{\tt L. M. Procopio; L. A. Rozema; B. Daki\'c; P. Walther}.
\newblock {Comment on Adler's ``Does the Peres experiment using photons test
  for hyper-complex (quaternionic) quantum theories?''}.
\newblock arXiv:1607.01648 [quant-ph] (2016).

\bibitem{Sobhani:2017yee}
{\tt H. Sobhani; H; Hassanabadi; W. S. Chung}.
\newblock {``Observations of the Ramsauer–Townsend effect in quaternionic
  quantum mechanics''}.
\newblock {\em Eur. Phys. J.}, {\bf C77}(6):425, (2017).

\bibitem{DeLeo:2019bcw}
{\tt S. De Leo; C A. A. de Souza; G. Ducati}.
\newblock {``Quaternionic perturbation theory''}.
\newblock {\em Eur. Phys. J. Plus}, {\bf 134}({\bf 3}):113, (2019).

\bibitem{Hassanabadi:2017wrt}
{\tt H. Hassanabadi; H. Sobhani; A. Banerjee}.
\newblock {``Relativistic scattering of fermions in quaternionic quantum
  mechanics''}.
\newblock {\em Eur. Phys. J.}, {\bf C77}(9):581, (2017).

\bibitem{Hassanabadi:2017jiz}
{\tt H. Hassanabadi, H. Sobhani; W. S. Chung}.
\newblock {``Scattering Study of Fermions Due to Double Dirac Delta Potential
  in Quaternionic Relativistic Quantum Mechanics''}.
\newblock {\em Adv. High Energy Phys.}, {\bf 2018}:8124073, 2018.

\bibitem{Giardino:2018lem}
{\tt S. Giardino}.
\newblock {``Non-anti-hermitian Quaternionic Quantum Mechanics''}.
\newblock {\em Adv. Appl. Clifford Algebras}, {\bf 28}(1):19, (2018).

\bibitem{Giardino:2022kxk}
{\tt S. Giardino}.
\newblock {``Quaternionic scalar field in the real Hilbert space''}.
\newblock {\em Int. J. Mod. Phys.}, {\bf A37}(15):2250101, (2022).

\bibitem{Giardino:2017yke}
{\tt S. Giardino}.
\newblock {``Quaternionic quantum particles''}.
\newblock {\em Adv. Appl. Clifford Algebras}, {\bf 29}(4):83, (2019).

\bibitem{Giardino:2017pqq}
{\tt S. Giardino}.
\newblock {``Quaternionic quantum particles: new solutions''}.
\newblock {\em Can. J. Phys.}, {\bf 99}:4, 6 (2017).

\bibitem{Giardino:2019xwm}
{\tt S. Giardino}.
\newblock {``Virial theorem and generalized momentum in quaternionic quantum
  mechanics''}.
\newblock {\em Eur. Phys. J. Plus}, {\bf 135}(1):114, (2020).

\bibitem{Giardino:2020cee}
{\tt S. Giardino}.
\newblock {``Square-well potential in quaternic quantum mechanics''}.
\newblock {\em Europhys. Lett.}, {\bf 132}:20007, 9 (2020).

\bibitem{Giardino:2021ofo}
{\tt S. Giardino}.
\newblock {``Quaternionic quantum harmonic oscillator''}.
\newblock {\em Eur. Phys. J. Plus}, {\bf 136}(1):120, (2021).

\bibitem{Giardino:2020ztf}
{\tt S. Giardino}.
\newblock {``Quaternionic elastic scattering''}.
\newblock {\em EPL}, {\bf 132}(5):50010, (2020).

\bibitem{Giardino:2021lov}
{\tt S. Giardino}.
\newblock {``Quaternionic Klein-Gordon equation''}.
\newblock {\em Eur. Phys. J. Plus}, {\bf 136}(6):612, (2021).

\bibitem{Giardino:2021mjj}
{\tt S. Giardino}.
\newblock {``Quaternionic Dirac free particle''}.
\newblock {\em Int. J. Mod. Phys.}, {\bf A36}(33):2150257, (2021).

\bibitem{Giardino:2022gqn}
{\tt S. Giardino}.
\newblock {``Quaternionic fermionic field''}.
\newblock {\it accept by Int. J. Mod. Phys. A} (2022).

\bibitem{Dirac:1937pqm}
\texttt{P. A. M. Dirac}.
\newblock {\em {``The principles of quantum mechanics''}}.
\newblock {\it Oxford U. P.} (1978).

\bibitem{Cahay:2019bqp}
{\tt M. Cahay; G. B. Purdy; D. Morris}.
\newblock {``On the quaternion representation of the Pauli spinor of an
  electron''}.
\newblock {\em Phys. Scripta}, {\bf 94}(8):085205, (2019).

\end{thebibliography}
\begin{footnotesize}

\end{footnotesize}

\end{document}